\documentclass[english,aps,superscriptaddress,showpacs]{revtex4}
\usepackage[T1]{fontenc}
\usepackage[utf8]{inputenc}
\usepackage{color}
\usepackage{babel}
\usepackage{amstext}
\usepackage{esint}
\usepackage[unicode=true, pdfusetitle,
 bookmarks=false,
 breaklinks=false,pdfborder={0 0 0},backref=false,colorlinks=true]{hyperref}
\usepackage{mathrsfs}

\hypersetup{
 citecolor=blue}
\makeatletter
\@ifundefined{textcolor}{}
{
 \definecolor{BLACK}{gray}{0}
 \definecolor{WHITE}{gray}{1}
 \definecolor{RED}{rgb}{1,0,0}
 \definecolor{GREEN}{rgb}{0,1,0}
 \definecolor{BLUE}{rgb}{0,0,1}
 \definecolor{CYAN}{cmyk}{1,0,0,0}
 \definecolor{MAGENTA}{cmyk}{0,1,0,0}
 \definecolor{YELLOW}{cmyk}{0,0,1,0}
 }
\makeatother

\begin{document}

\title{Parity violating quantum kinetic theory in (2+1)-dimensions}
\author{Jiunn-Wei Chen}
\affiliation{Department of Physics, National Center for Theoretical Sciences, and Leung
Center for Cosmology and Particle Astrophysics, National Taiwan University,
Taipei 10617, Taiwan}
\author{Jian-Hua Gao}
\affiliation{Shandong Provincial Key Laboratory of Optical Astronomy and
Solar-Terrestrial Environment, School of Space Science and Physics, Shandong
University at Weihai, Weihai 264209, China}
\affiliation{Key Laboratory of Quark and Lepton Physics (Central China Normal
University), Ministry of Education, China}
\author{Juan Liu}
\affiliation{Interdisciplinary Center for Theoretical Study and Department of Modern
Physics, University of Science and Technology of China, Hefei 230026, China}
\author{Shi Pu}
\affiliation{Department of Physics, National Center for Theoretical Sciences, and Leung
Center for Cosmology and Particle Astrophysics, National Taiwan University,
Taipei 10617, Taiwan}
\affiliation{Interdisciplinary Center for Theoretical Study and Department of Modern
Physics, University of Science and Technology of China, Hefei 230026, China}
\author{Qun Wang}
\affiliation{Interdisciplinary Center for Theoretical Study and Department of Modern
Physics, University of Science and Technology of China, Hefei 230026, China}

\begin{abstract}
We study the kinetic theory for a (2+1)-dimensional fermionic system with
special emphasis on the parity violating properties associated with the
fermion mass. The Wigner function approach is used to derive hydrodynamical
transport coefficients to the first spatial derivative order. As a first
attempt, the collisions between fermions are neglected. The resulting system
is dissipationless. The parity violating Hall electric conductivity has the
same temperature and chemical potential dependence as the quantum field
theory result at one-loop. Vorticity dependent transport properties, which
were not considered before, also emerge naturally in this approach.
\end{abstract}


\begin{flushright}
{\mbox{\hspace{10cm}}} USTC-ICTS-13-09
\end{flushright}

\maketitle

\section{Introduction}

\label{introduction}

Recently an asymmetry of certain charge-dependent azimuthal correlations
have been observed by the STAR \cite{Abelev2009,Abelev2010} and PHENIX
experiments \cite{PHENIX_charge} at the Relativistic Heavy-Ion Collider
(RHIC) and by the ALICE experiment \cite{Christakoglou2011} at the Large
Hadron Collider (LHC). Such an asymmetry disappears at low energies where
the chiral symmetry is broken \cite{Mohanty2011}. One possible explanation
of this phenomenon is the Chiral Magnetic Effect (CME) 
\cite{Kharzeev2008,Fukushima2008,Kharzeev2011}. The CME and related topics have
been studied in several approaches, including AdS/CFT correspondence 
\cite{Erdmenger2009,Banerjee2011,Torabian2009a,Rebhan2010,Kalaydzhyan:2011vx},
relativistic hydrodynamics 
\cite{Son2009,Pu:2010as,Sadofyev:2010pr,Kharzeev2011a}, kinetic theory 
\cite{Gao2012a,Son2012,Son2012a,Stephanov2012a,Chen2012,Pu2012}, lattice
calculations \cite{Abramczyk:2009gb,Buividovich:2009wi,Buividovich:2009zzb,Buividovich:2010tn,Yamamoto:2011gk}
and approaches from field theory
\cite{Fukushima2008,Metlitski2005,Newman2006,Charbonneau2010,Lublinsky2010,Asakawa2010,
Landsteiner:2011cp,Hou:2011ze,Golkar:2012kb,Jensen2012a,Jensen2013,Gorbar:2013upa,Huang2013}, 
for a recent review, see e.g. \cite{Kharzeev2013}.

Different from the (3+1)-dimensions [(3+1)D], the mass term in the Dirac
equation in (2+1)-dimensions [(2+1)D] explicitly breaks the parity 
\cite{Deser:1981wh}. After integration over the fermionic degrees of freedom, a
Chern-Simons (CS) term $\propto\epsilon^{\sigma\rho\alpha}A_{\sigma}
F_{\rho\alpha}$, with $A_{\sigma}$ an abelian gauge field, is induced in the
effective action $S_{eff}$ from the one loop correction. The fermion number
current $j_v^{\sigma}$ can then be obtained by taking the functional
derivative of $S_{eff}$ with respect to $A_{\sigma}$ 
\cite{Niemi1983,Redlich1984,Redlich1984a,Ishikawa1984,Ishikawa1985},
\begin{equation}
j_v^{\sigma}=\frac{i}{Q}\frac{\delta S_{eff}}{\delta A^{\sigma}} 
=-\mathrm{sign}(m)\frac{Q}{8\pi}\epsilon^{\sigma\rho\alpha}F_{\rho\alpha},
\label{eq:j_regulated_01}
\end{equation}
where $m$ and $Q$ are the mass and charge of the fermion respectively, 
and $\mathrm{sign}(x)$ is a sign function denoting the sign of $x$. Therefore,
the fermionic number density $j_v^{0}$ is modified by the magnetic field,
and the spatial components show a behavior of Hall conductivity 
$j^{1}\propto E^{2}$ and $j^{2}\propto E^{1}$. In the massless case, parity
is conserved classically but broken quantum mechanically (parity anomaly).
The one loop effective action has an ultraviolet divergence. To regularize
this divergence, one can use the standard Pauli-Villars regularization
method which preserves gauge symmetry,
\begin{equation}
S_{eff}^{reg}[m=0]=S_{eff}[m=0]-\lim_{M\rightarrow\infty}S_{eff}[M],
\end{equation}
where $M$ is a hypothetical mass and plays the role of a cutoff. However,
this cutoff term will induce a CS term just like in the massive case which
breaks parity \cite{Redlich1984,Redlich1984a}. In a non-Abelian gauge field
theory, the CS term proportional to $\epsilon^{\sigma\rho\alpha}\text{Tr }%
[A_{\sigma}\partial _{\rho}A_{\alpha}+\frac{2}{3}A_{\sigma}A_{\rho}A_{%
\alpha}]$ breaks parity, the prefactor of this non-abilean CS term needs to
be quantized to preserve gauge symmetry under a large gauge transformation
\cite{Dunne:1998qy}. There are subtle issues about how to preserve this
gauge invariance at non-abelian theories at finite temperatures. Here we
just work in the Abelian case and compare our results to quantum field
theory calculations at one-loop. At finite chemical potential $\mu$ and zero
temperature, if $\mu^{2}<m^{2}$, there is no dependence on $\mu$ \cite%
{Niemi1984,Niemi1986} and the current returns to Eq. (\ref{eq:j_regulated_01}%
). If $\mu^{2}>m^{2}$, the Chern-Simons term vanishes \cite{Sisakian:1997cp}%
. For recent reviews about the Chern-Simons theory, see, e.g., Ref. \cite%
{Niemi1986,Dunne:1998qy}.

Parity violating hydrodynamics in (2+1)D with Hall viscosity has recently
drawn a lot of attention in effective and holographic theories \cite%
{Nicolis:2011ey,Saremi:2011ab,Hoyos:2011ez,Chen:2011fs,Delsate:2011qp,Kimura:2011ef,Chen:2012ti}
as well as in condensed matter physics \cite%
{Read:2008rn,Avron:1995fg,acron98,PhysRevB.84.085316,PhysRevLett.107.075502}. 
A systematic discussion about the constraints from the second law of
thermodynamics in (2+1)D relativistic hydrodynamics is given by 
Ref. \cite{Jensen:2011xb}. Similar constraints can be derived in a curved space 
\cite{Banerjee:2012iz,Bhattacharyya:2012xi}. The equality constraints on (2+1)D
parity violating hydrodynamics are derived in Ref. \cite{Jensen2012} 
from general properties of Euclidean field theory in equilibria 
with slowly varying background metrics and gauge fields.

In previous works by some of us \cite{Gao2012a,Chen2012}, a quantum kinetic
theory was proposed to describe the CME and the Chiral Vortical Effect (CVE)
through the gauge invariant Wigner function \cite{Gao2012a}. The $U(1)$ and $%
U(1)_{A}$ currents induced by magnetic field and vorticity were obtained
from the vector and axial-vector components of the Wigner function. The
axial $U(1)_{A}$ current led to a local polarization effect along the
vorticity direction in peripheral heavy ion collisions. A chiral kinetic
equation was also derived from the Wigner function with features of the
Berry phase and monopole \cite{Chen2012}.

In this paper, we try to extend our previous works of quantum kinetic theory
in (3+1)D to (2+1)D. As a first attempt, we turn off interactions among
fermions and assume a constant electromagnetic background field $F^{\mu\nu}$
counted in the same order as the hydrodynamic scale. Then we expand the
equations of Wigner functions in powers of the Kundesn number $K$, defined
as the ratio of the mean free path to the macroscopic and hydrodynamic
scale, which is equivalent to gradient expansion. In the zeroth order, we
reproduce all macroscopic quantities as well as equations of motion in an
ideal fluid. In the first order, the electromagnetic field and vorticity
appear in the current and energy-momentum tensor. To test the
self-consistency of our approach, we compute constraints up to the second
order. After integrating over the 3-momenta, the energy-momentum tensor, the
fermion number current and the entropy current with parity violating terms
can be obtained. At the zero temperature and zero chemical potential, the
current induced by a magnetic field is consistent with Eq. (\ref%
{eq:j_regulated_01}) in Chern-Simons theory. The current induced by
vorticity is also obtained. We finally prove that the entropy is conserved.

The paper is organized as follows. in Sec. \ref{dirac}, we give basic
properties of Dirac matrices and parity for fermionic fields in (2+1)D. Sec. %
\ref{sec:Quantum-kinetic-approach} is devoted to quantum kinetic theory in
(2+1)D. In Sec. \ref{solve-w} we solve the Wigner function order by order in
the space-time derivative expansion. In Sec. \ref{current}, the
energy-momentum tensor, the fermion number and entropy current are obtained
by momentum integration. The Hall and vorticity term are reproduced. The
Landau frame is discussed in Sec. \ref{landau}. Finally we present a summary
and conclusion in Sec. \ref{sec:Discussion-and-conclusion}.

Our conventions and notations are: $g^{\mu\nu}=\text{diag}\{+,-,-\}$, $%
u^{\mu}u_{\mu}=1$, $\Delta^{\mu\nu}=g^{\mu\nu}-u^{\mu}u^{\nu}$, with $%
u^{\mu}(x)$ the fluid velocity. For a vector $h^{\mu}$, it can be decomposed
as $h^{\mu}=(u\cdot h)u^{\mu}+\bar{h}^{\mu}$, with $\bar{h}%
^{\mu}=\Delta^{\mu\lambda}h_{\lambda}$. We also define the comoving
derivative of a space-time quantity $a$ as $\dot{a}=da/dt=u^{\mu}\partial_{%
\mu}a$.

\section{Dirac $\protect\gamma$ matrices and parity in (2+1)D}

\label{dirac} We choose the following representation of the $\gamma$
matrices as follows,
\begin{eqnarray}
\gamma^{0}=\sigma_{2}=\left(%
\begin{array}{cc}
0 & -i \\
i & 0%
\end{array}%
\right), & \gamma^{1}=i\sigma_{1}=\left(%
\begin{array}{cc}
0 & i \\
i & 0%
\end{array}%
\right), & \gamma^{2}=i\sigma_{3}=\left(%
\begin{array}{cc}
i & 0 \\
0 & -i%
\end{array}%
\right),  \label{eq:gamma}
\end{eqnarray}
where $\sigma_{i}$ are Pauli matrices. The $\gamma$ matrices satisfy
\begin{eqnarray}
\{\gamma_{\alpha},\gamma_{\beta}\} & = &
2g_{\alpha\beta},\;[\gamma_{\alpha},\gamma_{\beta}]=2i\epsilon_{\alpha\beta%
\rho}\gamma^{\rho}  \nonumber \\
\gamma_{\alpha}\gamma_{\beta} & = &
g_{\alpha\beta}+i\epsilon_{\alpha\beta\rho}\gamma^{\rho}
\end{eqnarray}
where $\epsilon^{\alpha\beta\rho}$ is the Levi-Civita (anti-symmetric)
tensor with $\epsilon^{012}=\epsilon_{012}=1$. The Hermitian conjugation is $%
(\gamma^{\mu})^{\dagger}=\gamma^{0}\gamma^{\mu}\gamma^{0}$. Note that there
is no chirality in (2+1)D due to the absence of $\gamma^{5}$ since $%
i\gamma^{0}\gamma^{1}\gamma^{2}=-1$. Note that the Dirac gamma matrices in
Eq. (\ref{eq:gamma}) provide one of the irreducible representations. One can
immediately write down another irreducible representation by flipping the
sign of $\gamma^{\sigma}$. The nature of the lowest Landau level is
different for two representations, e.g. if $\gamma^{\sigma}$ corresponds to
the positive energy state $E_{0}=m$ then $-\gamma^{\sigma}$ corresponds to
the negative energy state $E_{0}=-m$. In real (2+1)D relativistic systems,
such as graphene, one uses a reducible representation which combines these
two irreducible representations (see, e.g. Ref. \cite{Shovkovy:2012zn}).

The parity transformation in (2+1)D is defined by flipping the sign of one
spatial component of a vector, for instance, $\mathbf{x}\rightarrow
\tilde{\mathbf{x}}$ where $\tilde{\mathbf{x}}=(-x^{1},x^{2})$, and then a spinor
transform as $\psi(t,\mathbf{x})\rightarrow\gamma^{1}
\psi(t,\tilde{\mathbf{x}})$. So we see that the mass term transform as 
$\bar{\psi}(t,\mathbf{x})\psi(t,\mathbf{x})\rightarrow-\bar{\psi}(t,\tilde{\mathbf{x}})
\psi(t,\tilde{\mathbf{x}})$, which is a pseudoscalar.

\section{Quantum kinetic equation for Wigner function}

\label{sec:Quantum-kinetic-approach} In a quantum kinetic approach, we
replace the phase-space distribution $f(x,p)$ by the Wigner function $W(x,p)$
with the space-time position $x$ and the (2+1)-momentum $p$. It is is
defined as the ensemble average of the gauge invariant Wigner operator 
\cite{Elze:1986qd,Vasak:1987um,Elze:1989un}. For fermions with mass $m$ and
charge $Q$, the $(\alpha,\beta)$ component of the Wigner function operator
can be written as
\begin{equation}
\hat{W}_{\alpha\beta}(x,p) = \int\frac{d^{3}y}{(2\pi)^{3}}e^{-ip\cdot y}
\bar{\psi}_{\beta}(x+y/2)U(x,y)\psi_{\alpha}(x-y/2),
\label{wigner}
\end{equation}
where the gauge link $U(x,y)$ ensures the gauge invariance of 
$\hat{W}_{\alpha\beta}$ and is defined by
\begin{equation}
U(x,y)\equiv\exp\left[-iQ\int_{x-y/2}^{x+y/2}dz^{\mu}A_{\mu}(z)\right].  
\label{link}
\end{equation}
As a first attempt, we consider a system of collisionless fermions in a
constant external electromagnetic field $F_{\mu\nu}$. In this case, we can
drop the path ordering in the gauge link in Eq. (\ref{link}).

In quantum field theory, physical quantities correspond to
matrix elements of operators under the time ordering $(\mathcal{T})$. This is
also true for the Wigner function. Using 
\begin{equation}
\mathcal{T}\hat{W}=:\hat{W}:+\langle 0|\mathcal{T}\hat{W}|0\rangle ,  
\label{eq:ordering}
\end{equation}
where the colons : : indicate a normal ordering, we can isolate the
medium effect which comes from the assemble average of $:\hat{W}:$
with a constant electromagnetic background, and the vacuum
(meaning the zero chemical potential and zero temperature ground state)
contribution $\langle 0|\mathcal{T}\hat{W}|0\rangle $.

When the vacuum contribution vanishes, one can simply use the normal
ordering part as was done in 
Ref. \cite{Elze:1986qd,Vasak:1987um,Elze:1989un,Gao2012a,Chen2012}. For example, in
(3+1)D, one can not write down a non-vanishing and gauge invariant expression
for the matrix element $\langle 0|\mathcal{T} \{ \text{Tr } [ \hat{W}\gamma ^{\mu }]\} |0\rangle $ 
using combinations of $F^{\mu \nu }$, $D_{\mu }$, $g^{\mu \nu }$, 
and $\epsilon ^{\mu \nu \alpha\beta }$. The simplest combination $\partial _{\mu }F^{\mu \nu }$ 
vanishes in vacuum. Analogously, one cannot write down a
non-vanishing contribution for the axial current 
$\langle 0|\mathcal{T} \{ \text{Tr } [ \hat{W}\gamma ^{\mu }\gamma ^{5} ] \} |0\rangle $ either. 
The simplest choice $\epsilon ^{\mu \nu \alpha \beta }\left( \partial _{\nu
}+iQA_{\nu }\right) F_{\alpha \beta }=iQ\epsilon ^{\mu \nu \alpha \beta
}A_{\nu }F_{\alpha \beta }$ is gauge dependent. Therefore, there
will be no vacuum contributions to these currents. This argument is
consistent with the fact that all CME and CVE coefficients vanish at zero
temperature, charge and chiral chemical potential limit. In (2+1)D, however,
we have a vacuum contribution to the vector current as shown in Eq. (1).
Thus, we need to associate the Wigner distribution function to the ensemble
average of $\mathcal{T}\hat{W}$.

From the equation of motion of a Dirac field, the master equation
for the Wigner operator is obtained \cite{Elze:1986qd,Vasak:1987um,Elze:1989un}
\begin{equation}
\gamma ^{\mu }\left( p^{\mu }+\frac{1}{2}i\nabla ^{\mu }-m\right) \mathcal{T}
\hat{W}(x,p)=0,
\end{equation}
where $\nabla ^{\mu }\equiv \partial _{x}^{\mu }-QF^{\mu \nu
}\partial _{\nu }^{p}$. The vacuum matrix element of this operator
equation implies $\langle 0|\mathcal{T}\hat{W}|0\rangle$ satisfies the same equation as
well. Then by eq. (\ref{eq:ordering}), $:\hat{W}:$ and its
ensemble average, $W\equiv \langle :\hat{W}:\rangle$, also satisfies the same
equation, 
\begin{equation}
\gamma ^{\mu }\left( p^{\mu }+\frac{1}{2}i\nabla ^{\mu }-m\right) W(x,p)=0.
\label{eq-c}
\end{equation} 

Although the medium effect can be derived from the kinetic theory
based on Eq.\ (\ref{eq-c}), this approach is not restrictive enough to fully
determine the vacuum contributions. Thus, we will just match them to the
quantum field theory results and combine them with the medium contributions
later. 

The Wigner function $W$ can be expanded in terms of $4$ independent
generators $\{1,\gamma^{\sigma}\}$ of the Clifford algebra,
\begin{equation}
W=\frac{1}{2}(\mathscr{F}+\gamma^{\sigma}\mathscr{V}_{\sigma}),
\label{eq:decomposition_01}
\end{equation}
where
\begin{eqnarray}
\mathscr{F} & \equiv & \text{Tr }[W],\;\mathscr{V^{\sigma}}\equiv\text{Tr }%
[\gamma^{\sigma}W].
\end{eqnarray}
We can obtain the medium part of the fermionic number current $j^{\sigma}$
from $\mathscr{V}^{\sigma}$ by integration over $p$,
\begin{equation}
j_m^{\sigma}=\int d^{3}p\mathscr{V}^{\sigma}=\int d^{3}p\text{Tr }%
[\gamma^{\sigma}W].  \label{eq:current_01}
\end{equation}
The total fermionic number current $j^{\sigma}$ is the sum of the medium and
vacuum part,
\begin{equation}
j^{\sigma}=j_m^{\sigma}+j_{v}^{\sigma},
\end{equation}
where $j_{v}^{\sigma}$ is given by Eq. (\ref{eq:j_regulated_01}). The
energy-momentum tensor can also be obtained from $\mathscr{V}^{\sigma}$,
\begin{equation}
T^{\sigma\rho}=\frac{1}{2}\int d^{3}pp^{(\sigma}\mathscr{V}^{\rho)}=\frac{1}{%
2}\int d^{3}pp^{(\sigma}\text{Tr }[\gamma^{\rho)}W],
\label{eq:energy_momentum_tensor}
\end{equation}
where the parentheses denote index symmetrization.

Substituting Eq. (\ref{eq:decomposition_01}) into Eq. (\ref{eq-c}) yields,
\begin{eqnarray}
p\cdot\mathscr{V}-m\mathscr{F} & = & 0,  \label{eq:dirac_eq_01} \\
\nabla\cdot\mathscr{V} & = & 0,  \label{eq:dirac_eq_02} \\
p^{\sigma}\mathscr{F}-\frac{1}{2}\epsilon^{\sigma\lambda\rho}\nabla_{\lambda}%
\mathscr{V}_{\rho}-m\mathscr{V}^{\sigma} & = & 0,  \label{eq:dirac_eq_03} \\
\frac{1}{2}\nabla^{\sigma}\mathscr{F}+\epsilon^{\sigma\lambda\rho}p_{\lambda}%
\mathscr{V}_{\rho} & = & 0.  \label{eq:dirac_eq_04}
\end{eqnarray}
In Eqs. (\ref{eq:dirac_eq_01}-\ref{eq:dirac_eq_04}), there are $8$ highly
consistent equations for 4 components of $\mathscr{F}$ and $\mathscr{V}%
^{\sigma}$.

\section{Solving Wigner function in expansion of space-time derivatives}

\label{solve-w} Since we are interested in long wave-length physics, we set
up the system near equilibrium so that we can expand $\mathscr{F}$ and $%
\mathscr{V}^{\sigma}$ in powers of the space-time derivative $\partial_{x}$.
We also assume that the background field $F^{\rho\sigma}=\partial^{\rho}A^{%
\sigma}-\partial^{\sigma}A^{\rho}$ is of the same order as $\partial_{x}$.
In this case, $\mathscr{V}^{\sigma}$ and $\mathscr{F}$ can be written as,
\begin{eqnarray}
\mathscr{V}^{\sigma} & = & \mathscr{V}_{(0)}^{\sigma}+\mathscr{V}%
_{(1)}^{\sigma}+O(\partial_{x}^{2}),  \label{eq:expansion_01} \\
\mathscr{F} & = & \mathscr{F}_{(0)}+\mathscr{F}_{(1)}+O(\partial_{x}^{2}),
\label{eq:expansion_02}
\end{eqnarray}
where the indices $(0),(1)$ denote the zeroth and the first order
respectively. From Eq. (\ref{eq:dirac_eq_03},\ref{eq:dirac_eq_04}), we see
that $\mathscr{V}_{(n)}^{\sigma}$ is related to $\mathscr{F}_{(n-1)}$ and
that $\mathscr{F}_{(n)}$ is related to $\mathscr{V}_{(n-1)}^{\sigma}$. So we
can solve $\mathscr{V}^{\sigma}$ and $\mathscr{F}$ order by order.

\subsection{Zeroth order}

\label{sub:Zeroth-order} At the zeroth order, Eqs.(\ref{eq:dirac_eq_01}-\ref%
{eq:dirac_eq_04}) become
\begin{eqnarray}
p\cdot\mathscr{V}_{(0)}-m\mathscr{F}_{(0)} & = & 0,  \label{eq:dirac_eq_01-1}
\\
p^{\sigma}\mathscr{F}_{(0)}-m\mathscr{V}_{(0)}^{\sigma} & = & 0,
\label{eq:dirac_eq_03-1} \\
\epsilon^{\sigma\lambda\rho}p_{\lambda}\mathscr{V}_{\rho}^{(0)} & = & 0.
\label{eq:dirac_eq_04-1}
\end{eqnarray}
The general forms for $\mathscr{V}_{(0)}^{\mu}$ and $\mathscr{F}$ satisfying
the above equations are
\begin{eqnarray}
\mathscr{V}_{(0)}^{\sigma} & = & p^{\sigma}V\delta(p^{2}-m^{2}),
\label{eq:vv_sol_01} \\
\mathscr{F}_{(0)} & = & mV\delta(p^{2}-m^{2}),  \label{eq:ff_sol_01}
\end{eqnarray}
where $V$ is a function of $x$ and $p$.

As we mentioned before, we set up a perturbative scheme around the
equilibrium state. Therefore, the zeroth order should correspond to an
equilibrium non-interacting ideal gas, where the macroscopic quantities can
be obtained by the Fermi-Dirac distribution. On the other hand, these
quantities can also be obtained from $\mathscr{V}_{(0)}^{\sigma}$ or $%
\mathscr{F}_{(0)}$. We set $V$ to be
\begin{equation}
V=\frac{1}{2\pi^{2}}\sum_{e=\pm}\frac{1}{e^{e(u\cdot p-\mu)/T}+1}%
\theta(eu\cdot p-|m|),  \label{eq:V_01}
\end{equation}
where $e=\pm$ denote fermions/anti-fermions, $u^{\sigma}$ denotes the fluid
velocity, $T$ is the temperature, $\mu$ is the chemical potential, and $%
\theta(x)=1,0$ for positive/negative $x$ with $\theta(0)=1/2$. Note that $V$
becomes a space-time dependent function via the dependence on $\mu(x)$, $T(x)
$ and $u^{\sigma}(x)$. Integrating $\mathscr{V}_{(0)}^{\sigma}$ over the
(2+1) momenta, we obtian the current at the leading order,
\begin{eqnarray}
j_{m(0)}^{\sigma} & = & u^{\sigma}\frac{1}{(2\pi)^{2}}\int d^{2}p\left[\frac{%
1}{e^{\beta(E_{p}-\mu)}+1}-\frac{1}{e^{\beta(E_{p}+\mu)}+1}\right],
\end{eqnarray}
which is just $n_{0}u^{\sigma}$.

\subsection{First order}

At the first order, Eqs. (\ref{eq:dirac_eq_01}-\ref{eq:dirac_eq_04}) becomes
\begin{eqnarray}
p\cdot\mathscr{V}_{(1)}-m\mathscr{F}_{(1)} & = & 0,  \label{eq:dirac_eq_01-2}
\\
\nabla\cdot\mathscr{V}_{(0)} & = & 0,  \label{eq:dirac_eq_02-2} \\
p^{\sigma}\mathscr{F}_{(1)}-\frac{1}{2}\epsilon^{\sigma\lambda\rho}\nabla_{%
\lambda}\mathscr{V}_{\rho}^{(0)}-m\mathscr{V}_{(1)}^{\sigma} & = & 0,
\label{eq:dirac_eq_03-2} \\
\frac{1}{2}\nabla^{\sigma}\mathscr{F}_{(0)}+\epsilon^{\sigma\lambda\rho}p_{%
\lambda}\mathscr{V}_{\rho}^{(1)} & = & 0.  \label{eq:dirac_eq_04-2}
\end{eqnarray}
Given Eqs. (\ref{eq:vv_sol_01},\ref{eq:ff_sol_01},\ref{eq:dirac_eq_01-2},\ref%
{eq:dirac_eq_03-2}), it can be verified that Eq. (\ref{eq:dirac_eq_04-2})
holds automatically provided Eq. (\ref{eq:dirac_eq_02-2}) is satisfied. So
we see that Eq. (\ref{eq:dirac_eq_02-2}) is the basic equation for the
zero-th order solution of $\mathscr{V}_{(0)}$ and $\mathscr{F}_{(0)}$ which
provides constraints for $\mu,T,u^{\sigma}$. Inserting Eqs. (\ref%
{eq:vv_sol_01}, \ref{eq:V_01}) into Eq.(\ref{eq:dirac_eq_02-2}), we obtain,
\begin{eqnarray}
0 & = & \nabla_{\sigma}\mathscr{V}_{(0)}^{\sigma}  \nonumber \\
& = & \delta(p^{2}-m^{2})V_{u\cdot p}^{\prime}\times\left\{ \bar{p}^{2}\left[%
-Tu^{\sigma}\partial_{\sigma}\beta+\frac{1}{2}(\partial\cdot u)\right]\right.
\nonumber \\
& & -T\bar{p}^{\sigma}\left[\partial_{\sigma}(\beta\mu)+\beta QE_{\sigma}%
\right]+\left(\bar{p}^{\alpha}\bar{p}^{\lambda}-\frac{1}{2}\bar{p}%
^{2}\Delta^{\alpha\lambda}\right)\partial_{\langle\alpha}u_{\lambda\rangle}
\nonumber \\
& & \left.+m^{2}Tu^{\sigma}\partial_{\sigma}\beta-(u\cdot
p)Tu^{\sigma}\partial_{\sigma}(\beta\mu)+(u\cdot p)\bar{p}^{\sigma}\left[%
T\partial_{\sigma}\beta+\dot{u}_{\sigma}\right]\right\} .
\label{eq:constraint_01}
\end{eqnarray}
Here $V_{u\cdot p}^{\prime}$ denotes the derivative of $V$ with respect to $%
(u\cdot p)$ which does not include the derivative of $\theta$-function, so
we have
\begin{equation}
V_{u\cdot p}^{\prime}\equiv\left.\frac{\partial V}{\partial(u\cdot p)}%
\right|_{\bar{\theta}}=\frac{\beta}{u\cdot p}\frac{\partial V}{\partial\beta}%
=-\beta\frac{\partial V}{\partial(\beta\mu)}.
\end{equation}
We used the following notations
\begin{eqnarray}
& & \beta=1/T,\;\bar{p}^{\sigma}=p^{\sigma}-u^{\sigma}(u\cdot
p)=\Delta^{\sigma\rho}p_{\rho}  \nonumber \\
& & \partial_{\langle\alpha}u_{\lambda\rangle }=\frac{1}{2}%
\Delta_{\alpha\beta}\Delta_{\lambda\rho}(\partial^{\beta}u^{\rho}+\partial^{%
\rho}u^{\beta}-\Delta^{\beta\rho}\partial\cdot u)
\end{eqnarray}
where the angular brackets $\langle\rangle$ denote the traceless symmetrized
tensors. We also define $E^{\mu}$ and $B^{\mu}$ as
\begin{equation}
E^{\mu}=F^{\mu\nu}u_{\nu},\: B^{\mu}=\frac{1}{2}\epsilon^{\mu\nu\alpha}F_{%
\nu\alpha} =(-B,-E^{2},E^{1}).  \label{eq:E_B_def_01}
\end{equation}
Note that the magnetic field $B=\partial_{1}A^{2}-\partial_{2}A^{1}$ in
(2+1)D is a pseudo-scalar under rotation and the spatial components of $%
B^{\mu}$ are the Hall electric fields.

In order for Eq. (\ref{eq:constraint_01}) to be satisfied for any $p$ (this
condition will be relaxed after including collisions), we see that the
following conditions must be fulfilled
\begin{eqnarray}
\partial_{<\alpha}u_{\lambda>}=0, & & \partial_{\sigma}(\beta\mu)+\beta
QE_{\sigma}=0,  \nonumber \\
\dot{u}_{\sigma}-\partial_{\sigma}\ln T=0, & & \dot{T}=0,  \nonumber \\
u^{\sigma}\partial_{\sigma}(\beta\mu)=0, & & \partial\cdot u=0,
\label{eq:constraint_cond_01}
\end{eqnarray}
where we used the notation $\dot{X}=u^{\sigma}\partial_{\sigma}X$. The
conditions $\partial\cdot u=0$, $\partial_{<\alpha}u_{\lambda>}=0$ and $%
\partial_{\sigma}(\beta\mu)+\beta QE_{\sigma}=0$ imply that we neglect the
bulk viscous pressure, shear viscous tensor and heat conducting flow
respectively, so the system is dissipationless.

Contracting Eq. (\ref{eq:dirac_eq_03-2}) with $p_{\sigma}$ and using Eq. (%
\ref{eq:dirac_eq_01-2}) we obtian
\begin{equation}
\mathscr{F}_{(1)}=\frac{1}{2}Q\epsilon^{\rho\sigma\xi}p_{\rho}F_{\sigma\xi}V%
\delta^{\prime}(p^{2}-m^{2})+\frac{1}{2}\hat{G}\delta(p^{2}-m^{2}),
\label{eq:1st_F_01}
\end{equation}
where $\hat{G}$ is a function of $x$ and $p$ and will be determined later.
From Eqs. (\ref{eq:vv_sol_01},\ref{eq:dirac_eq_03-2},\ref{eq:1st_F_01}), we
have
\begin{eqnarray}
\mathscr{V}_{(1)}^{\rho} & = & \frac{1}{m}p^{\rho}\mathscr{F}_{(1)}-\frac{1}{%
2m}\epsilon^{\rho\sigma\xi}\nabla_{\sigma}\mathscr{V}_{\xi}^{(0)}  \nonumber
\\
& = & -\frac{1}{2m}\delta(p^{2}-m^{2})V_{u\cdot p}^{\prime}\left\{ p^{\rho}%
\left[\frac{1}{2}(u\cdot p)(u\cdot\omega)-(p\cdot\omega)\right]-\frac{1}{2}%
(u\cdot\omega)m^{2}u^{\rho}+m^{2}\omega^{\rho}\right\}  \nonumber \\
& & +QmV\delta^{\prime}(p^{2}-m^{2})B^{\rho}+\hat{G}\frac{1}{2m}%
p^{\rho}\delta(p^{2}-m^{2})+Q\frac{1}{2m}\bar{B}^{\rho}(u\cdot
p)C\delta(p^{2}-m^{2}),  \label{eq:V_1_temp_01}
\end{eqnarray}
where $C$ comes from the $(u\cdot p)$ derivative acting on the $\theta$%
-functions in $V$ of Eq. (\ref{eq:V_01})
\begin{eqnarray}
C & = & \frac{1}{2\pi^{2}}\sum_{e=\pm}\frac{e}{e^{\beta e(u\cdot p-\mu)}+1}%
\delta(eu\cdot p-|m|),  \label{eq:def_C}
\end{eqnarray}
where we have used the definition of the vorticity $\omega^{\rho}=\epsilon^{%
\rho\sigma\xi}\partial_{\sigma}u_{\xi}$. We see that the vorticity and
magnetic field emerge automatically in $\mathscr{V}^{\mu}$. They will
contribute to the current and energy-momentum tensor.

For simplicity, we can rewrite the unknown function $\hat{G}$ in Eqs. (\ref%
{eq:1st_F_01},\ref{eq:V_1_temp_01}) as
\begin{eqnarray}
\hat{G} & = & G+V_{u\cdot p}^{\prime}\left[\frac{1}{2}(u\cdot
p)(u\cdot\omega)-(p\cdot\omega)\right]+QC(u\cdot B).  \label{eq:gauge_01}
\end{eqnarray}
Then $\mathscr{V}_{(1)}^{\mu}$ and $\mathscr{F}_{(1)}$ become
\begin{eqnarray}
\mathscr{V}_{(1)}^{\rho} & = & -\frac{m}{2}\delta(p^{2}-m^{2})V_{u\cdot
p}^{\prime}\left[-\frac{1}{2}(u\cdot\omega)u^{\rho}+\omega^{\rho}\right]
\nonumber \\
& & +QmV\delta^{\prime}(p^{2}-m^{2})B^{\rho}+Q\frac{1}{2m}B^{\rho}(u\cdot
p)C\delta(p^{2}-m^{2})+G\frac{1}{2m}p^{\rho}\delta(p^{2}-m^{2}),
\label{eq:V_1_temp_02} \\
\mathscr{F}_{(1)} & = & Q(p\cdot B)V\delta^{\prime}(p^{2}-m^{2})+Q\frac{1}{2}%
(u\cdot B)C\delta(p^{2}-m^{2})  \nonumber \\
& & -\frac{1}{2}\delta(p^{2}-m^{2})V_{u\cdot p}^{\prime}\left[-\frac{1}{2}%
(u\cdot p)(u\cdot\omega)+(p\cdot\omega)\right]+\frac{G}{2}%
\delta(p^{2}-m^{2}),  \label{eq:F_01}
\end{eqnarray}
where we have used the fact that the combination of $C$ and $%
\delta(p^{2}-m^{2})$ leads to $\bar{p}^{\sigma}=0$, since $%
C\propto\delta(eu\cdot p-|m|)$.

\subsection{Second order}

In order to determine $G$, we have to consider the second order constraints
for $\mathscr{V}_{(1)}^{\mu}$. At the second order, Eqs. (\ref%
{eq:dirac_eq_01}-\ref{eq:dirac_eq_04}) become
\begin{eqnarray}
p\cdot\mathscr{V}_{(2)}-m\mathscr{F}_{(2)} & = & 0,
\label{eq:dirac_eq_01-2-1} \\
\nabla\cdot\mathscr{V}_{(1)} & = & 0,  \label{eq:dirac_eq_02-2-2} \\
p^{\rho}\mathscr{F}_{(2)}-\frac{1}{2}\epsilon^{\rho\sigma\xi}\nabla_{\sigma}%
\mathscr{V}_{\xi}^{(1)}-m\mathscr{V}_{(2)}^{\rho} & = & 0,
\label{eq:dirac_eq_03-2-1} \\
\frac{1}{2}\nabla^{\rho}\mathscr{F}_{(1)}+\epsilon^{\rho\sigma\xi}p_{\sigma}%
\mathscr{V}_{\xi}^{(2)} & = & 0.  \label{eq:dirac_eq_04-2-1}
\end{eqnarray}
The first constraint is provided by Eq. (\ref{eq:dirac_eq_02-2-2}).
Substituting Eq. (\ref{eq:dirac_eq_03-2-1}) into Eq. (\ref%
{eq:dirac_eq_04-2-1}) we get
\begin{equation}
m\nabla^{\rho}\mathscr{F}_{(1)}=\epsilon^{\rho\sigma\xi}p_{\sigma}\epsilon_{%
\xi\alpha\beta}\nabla^{\alpha}\mathscr{V}_{(1)}^{\beta}=p_{\sigma}[\nabla^{%
\rho}\mathscr{V}_{(1)}^{\sigma}-\nabla^{\sigma}\mathscr{V}_{(1)}^{\rho}],
\label{eq:constraint_02}
\end{equation}
which gives the second constraint.

Substituting the solutions (\ref{eq:V_1_temp_02}) and (\ref{eq:F_01}) into
constraint Eqs. (\ref{eq:dirac_eq_02-2-2}, \ref{eq:constraint_02}) and using
identities in Appendix \ref{sec:Appendix-A-Using}, we arrive at
\begin{equation}
\nabla_{\mu}\left[p^{\mu}G\delta(p^{2}-m^{2})\right]=0.
\label{eq:constraint_01-1}
\end{equation}
We can show that $G=0$ is a solution under certain physical
constraints (see Appedix \ref{sec:G=0}), so we obtain
\begin{eqnarray}
\mathscr{V}_{(1)}^{\rho} & = & -\frac{m}{2}V_{u\cdot p}^{\prime}\left[-\frac{%
1}{2}(u\cdot\omega)u^{\rho}+\omega^{\rho}\right]\delta(p^{2}-m^{2})+QmV%
\delta^{\prime}(p^{2}-m^{2})B^{\rho}  \nonumber \\
& & +\frac{1}{2m}QB^{\rho}(u\cdot p)C\delta(p^{2}-m^{2}).
\label{eq:V_1_temp_02-1}
\end{eqnarray}

\section{Currents and conservation laws}

\label{current} The medium part of the fermion number current can be
obtained by integrating over $p$ as in Eq. (\ref{eq:current_01}), combining
with the vacuum part, the total current is then
\begin{eqnarray}
j^{\sigma} & = & \left[n+\frac{1}{2}\xi(u\cdot\omega)+\xi_{B}(u\cdot B)%
\right]u^{\sigma}+\xi\bar{\omega}^{\sigma}+\xi_{B}\bar{B}^{\sigma},
\label{eq:current_02}
\end{eqnarray}
where the fermion number density $n$ and two coefficients $\xi$ and $\xi_{B}$
are given by
\begin{eqnarray}
n & = & \frac{1}{2\pi}\int_{|m|}^{\infty}dE_{p}E_{p}\left[\frac{1}{%
e^{\beta(E_{p}-\mu)}+1}-\frac{1}{e^{\beta(E_{p}+\mu)}+1}\right],  \nonumber
\\
\xi & = & mc_{-}(m),  \nonumber \\
\xi_{B} & = & Q\mathrm{sign}(m)\left[c_{+}(m)-\frac{1}{4\pi}\right],
\end{eqnarray}
and $c_{\pm}(m)$ is defined by
\begin{equation}
c_{\pm}(m)=\frac{1}{4\pi}\left[\frac{1}{e^{\beta(|m|-\mu)}+1}\pm\frac{1}{%
e^{\beta(|m|+\mu)}+1}\right].
\end{equation}
First we look at the magnetic field part of the current. At the zero
temperature limit $T\rightarrow0$, we have
\begin{eqnarray}
\xi_{B} & = & - \frac{Q}{4\pi}\mathrm{sign}(m)\theta(m^{2}-\mu^{2}).
\label{eq:medium-xi-b}
\end{eqnarray}
which is consistent to the results of Ref. \cite{Sisakian:1997cp,Niemi1986}.
Now we look at the vorticity part of the current, in the $T\rightarrow0$
limit, we have
\begin{eqnarray}
\xi & = & \frac{1}{4\pi}\mathrm{sign}(\mu)m\theta(\mu^{2}-m^{2}).
\end{eqnarray}
Note that there is no vacuum contribution to $\xi$ due to the symmetry
property of $j^{\sigma}$. The reason is as follows. The vacuum contribution,
if there is any, must be in the form of $C^{\prime}m$, where $C^{\prime}$ is
a constant independent of $\mu$, $T$ and $Q$, because under parity
transformation $m$ and $\xi$ transform as $m\rightarrow-m$ and $%
\xi\rightarrow-\xi$. The current $j^{\sigma}$ is also odd under charge
conjugation transformation $\mu\rightarrow-\mu$ and $Q\rightarrow-Q$.
However the vacuum contribution does not change sign under charge
conjugation transformation, hence is not allowed. Using identities in
Appendix \ref{sec:Appendix-A-Using} and Eq. (\ref{eq:constraint_cond_01}),
one can verify the fermion number conservation
\begin{equation}
\partial_{\sigma}j^{\sigma}=0.
\end{equation}
This is different from the case in (3+1)D where the conservation is broken
by anomaly.

By using Eq. (\ref{eq:energy_momentum_tensor}), the energy-momentum tensor
can be evaluated as,
\begin{eqnarray}
T^{\rho\sigma} & = &
u^{\rho}u^{\sigma}[\varepsilon+\kappa(u\cdot\omega)+2\kappa_{B}(u\cdot
B)]-\Delta^{\rho\sigma}P  \nonumber \\
& & +\kappa(u^{\rho}\bar{\omega}^{\sigma}+u^{\sigma}\bar{\omega}%
^{\rho})+\kappa_{B}(u^{\rho}\bar{B}^{\sigma}+u^{\sigma}\bar{B}^{\rho}),
\label{eq:stress-tensor}
\end{eqnarray}
where the energy density $\varepsilon$, the pressure $P$ and two
coefficients $\kappa$ and $\kappa_{B}$ are given by
\begin{eqnarray}
\varepsilon & = & \frac{1}{2\pi}\int_{|m|}^{\infty}dE_{p}E_{p}^{2}\left[%
\frac{1}{e^{\beta(E_{p}-\mu)}+1}+\frac{1}{e^{\beta(E_{p}+\mu)}+1}\right],
\nonumber \\
P & = & \frac{1}{4\pi}\int_{|m|}^{\infty}dE_{p}(E_{p}^{2}-m^{2})\left[\frac{1%
}{e^{\beta(E_{p}-\mu)}+1}+\frac{1}{e^{\beta(E_{p}+\mu)}+1}\right],  \nonumber
\\
\kappa & = & \frac{1}{8\pi}m\beta\int_{|m|}^{\infty}dE_{p}E_{p}\left[\frac{%
e^{\beta(E_{p}-\mu)}}{(e^{\beta(E_{p}-\mu)}+1)^{2}}+\frac{%
e^{\beta(E_{p}+\mu)}}{(e^{\beta(E_{p}+\mu)}+1)^{2}}\right]  \nonumber \\
& = & -\frac{1}{4m}\beta\frac{\partial}{\partial\beta}(\varepsilon-2P),
\nonumber \\
\kappa_{B} & = & \frac{1}{2}mQc_{-}(m).
\end{eqnarray}
Using identities in Appendix \ref{sec:Appendix-A-Using}, Eq. (\ref%
{eq:constraint_cond_01}), $\beta\partial P/\partial\beta=-(\varepsilon+P)$
and $\beta\partial P/\partial(\beta\mu)=n$, we can verify the
energy-momentum conservation
\begin{equation}
\partial_{\mu}T^{\mu\nu}=QF^{\nu\lambda}j_{\lambda}.
\end{equation}

The entropy current \cite{Israel:1979wp} is defined as
\begin{eqnarray}
s^{\sigma} & = & \beta(Pu^{\sigma}-\mu
j^{\sigma}+u_{\lambda}T^{\lambda\sigma})  \nonumber \\
& = & su^{\sigma}+\beta[(\kappa-\frac{1}{2}\mu\xi)(u\cdot\omega)+(2%
\kappa_{B}-\mu\xi_{B})(u\cdot B)]u^{\sigma}  \nonumber \\
& & +\beta(\kappa-\mu\xi)\bar{\omega}^{\sigma}+\beta(\kappa_{B}-\mu\xi_{B})%
\bar{B}^{\sigma},  \label{eq:entropy-current}
\end{eqnarray}
where we have used the Gibbs-Duhem relation $sT=P+\varepsilon-\mu n$.
Similarly we can confirm the entropy conservation,
\begin{equation}
\partial_{\sigma}s^{\sigma}=0,  \label{eq:entropy-cons}
\end{equation}
which means the system is non-dissipative.

\section{Landau frame}

\label{landau} The current and energy-momentum tensor Eqs. (\ref%
{eq:current_02},\ref{eq:stress-tensor}) are in a general frame. We can
transform them to the Landau frame. To this end we notice that the densities
of effective energy, fermion number and entropy are
\begin{eqnarray}
\varepsilon_{E} & = &
u_{\rho}u_{\sigma}T^{\rho\sigma}=\varepsilon+\kappa(u\cdot\omega)+2%
\kappa_{B}(u\cdot B),  \nonumber \\
n_{E} & = & u_{\rho}j^{\rho}=n+\frac{1}{2}\xi(u\cdot\omega)+\xi_{B}(u\cdot
B),  \nonumber \\
s_{E} & = & u_{\rho}s^{\rho}=P+\varepsilon_{E}-n_{E}\mu.
\end{eqnarray}
This is different from Isreal-Stwart theory \cite{Israel:1979wp} where the
energy and number density are corrected up to the second order when changing
the frame. However, as we mentioned in the introduction, the number density
is modified by the magnetic field and this is one of the properties in
(2+1)D QED. So it is expected that the energy density is also modified by
the magnetic field.

The Landau frame is defined as $\Delta_{\mu\nu}T^{\nu\alpha}u_{\alpha}=0$,
which define a new fluid velocity
\[
u_{E}^{\mu}=u^{\mu}+\frac{\kappa}{\epsilon+P}\bar{\omega}^{\mu}+\frac{%
\kappa_{B}}{\epsilon+P}\bar{B}^{\mu},
\]
then $T^{\rho\sigma}$ is written as
\begin{eqnarray}
T^{\rho\sigma} & = &
u_{E}^{\rho}u_{E}^{\sigma}(\varepsilon_{E}+P)-g^{\rho\sigma}P,  \nonumber \\
j^{\sigma} & = & n_{E}u_{E}^{\sigma}+\xi^{E}\bar{\omega}^{\sigma}+\xi_{B}^{E}%
\bar{B}^{\sigma},  \nonumber \\
s^{\sigma} & = & s_{E}u_{E}^{\sigma}+\xi_{s}^{E}\bar{\omega}%
^{\sigma}+\xi_{sB}^{E}\bar{B}^{\sigma},
\end{eqnarray}
with the new coefficients
\begin{eqnarray}
\xi^{E} & = & \xi-\frac{n}{\epsilon+P}\kappa,  \nonumber \\
\xi_{B}^{E} & = & \xi_{B}-\frac{n}{\epsilon+P}\kappa_{B},  \nonumber \\
\xi_{s}^{E} & = & \beta(\kappa-\mu\xi)-\frac{s}{\epsilon+P}\kappa,  \nonumber
\\
\xi_{sB}^{E} & = & \beta(\kappa_{B}-\mu\xi_{B})-\frac{s}{\epsilon+P}%
\kappa_{B}.
\end{eqnarray}

It is interesting to compare our results Eqs. (\ref{eq:current_02},\ref%
{eq:stress-tensor}) with the entropy principle analysis of Ref. \cite%
{Jensen:2011xb}. We note that the $\tilde{\chi}_{T}$ and $\tilde{\chi}_{E}$
terms in Ref. \cite{Jensen:2011xb} are actually our $\xi$ and $\xi_{B}$
terms respectively since $\tilde{E}_{\sigma}=-\bar{B}_{\sigma}$ and $%
\epsilon_{\eta\lambda\xi}u^{\lambda}\partial^{\xi}T=T\bar{\omega}_{\eta}$
with identities in Appendix \ref{sec:Appendix-A-Using} and constraints (\ref%
{eq:constraint_cond_01}). Note that we do not have shear and Hall viscosity
terms ($\eta$ and $\tilde{\eta}$ terms) which are dissipative since we have
no collision and dissipation as demonstrated by the entropy conservation in
Eq. (\ref{eq:entropy-cons}). Another dissipative term, the electric
conductivity term ($\sigma$ term) in Ref. \cite{Jensen:2011xb}, is also
absent in our approximation.

\section{Summary and conclusion \label{sec:Discussion-and-conclusion}}

We derive the parity violating fluid-dynamics of a fermionic system in
(2+1)-dimensions in quantum kinetic theory with the Wigner function. Using a
perturbative method in powers of the space-time derivative and
electromagnetic field, we determine the Wigner function to the first order
by solving a system of equations for the Wigner function to the second
order. Our main results are Eqs. (\ref{eq:current_02},\ref{eq:stress-tensor},%
\ref{eq:entropy-current}). In the zeroth order, the Wigner function gives
rise to the fermionic number current, the entropy current and the
energy-momentum tensor of an ideal gas. In order for the first order
equations to be satisfied, the constraints on the thermal variables are
imposed. We then solve the Wigner function up to the first order constrained
by the second order equations. Integrating over the energy-momentum for the
Wigner function one obtains the fermionic number current, the entropy
current and the energy-momentum tensor, where vorticity as well as
electromagnetic field terms appear naturally. At zero temperature, the Hall
conductivity is consistent with the previous result from quantum field
theory. We also prove the conservation of entropy which indicates that the
system is dissipationless.

Acknowledgement. QW thanks helpful discussion with Andreas Schmitt and Igor
Shovkovy. This work is supported by the NSFC under grant No. 11125524,
11105137, 11205150. JWC and SP are supported by the
NSC(99-2112-M-002-010-MY3) of ROC and CASTS \& CTS of NTU. SP will thank
Tomas Brauner and Sergej Moroz for the helpful discussion at the beginning
of this work. JHG is supported in part by CCNU-QLPL Innovation Fund
QLPL2011P01.

\appendix

\section{Useful identities}

\label{sec:Appendix-A-Using} 
In this appendix we give identities involving 
$u_{\alpha}$, $\partial_{\alpha}u_{\beta}$, $\omega_{\alpha}$, $B_{\alpha}$
and $E_{\alpha}$ under the conditions of Eq. (\ref{eq:constraint_cond_01}).
These identities are useful to verify Eq. (\ref{eq:dirac_eq_02-2-2}).

First we list main identities concerning $u_{\alpha}$, 
$\partial_{\alpha}u_{\beta}$, $\omega_{\alpha}$. From $\dot{u}^{\alpha}=\partial^{\alpha}\ln T$
and $\partial_{\sigma}\omega^{\sigma}=0$, we obtain
\begin{equation}
u_{\beta}\partial^{\beta}\omega_{\rho}=\omega_{\beta}\partial^{\beta}
u_{\rho}.  
\label{eq:omega-u}
\end{equation}
Then we have
\begin{equation}
u\cdot\partial(u\cdot\omega)=\dot{u}^{\sigma}\omega_{\sigma}=0.
\end{equation}
We can derive
\begin{eqnarray}
\partial_{\rho}u_{\sigma} & = & \frac{1}{2}(\epsilon_{\rho\sigma\tau}
\omega^{\tau}+u_{\sigma}\dot{u}_{\rho}+u_{\rho}\dot{u}_{\sigma})  
\nonumber \\
& = & \frac{1}{2}(u\cdot\omega)\epsilon_{\rho\sigma\tau}u^{\tau}
+u_{\rho}\dot{u}_{\sigma},  \nonumber \\
\partial_{\sigma}\omega^{\lambda} & = & \omega^{\lambda}\dot{u}_{\sigma}
+\frac{1}{2}(u\cdot\omega)u^{\lambda}\dot{u}_{\sigma}
+\frac{1}{2}(u\cdot\omega)\partial_{\sigma}u^{\lambda},  
\label{eq:du-do}
\end{eqnarray}
where we have used $\partial_{<\alpha}u_{\lambda>}=0$ in 
Eq. (\ref{eq:constraint_cond_01}) and Eq. (\ref{eq:omega-u}). 
Using Eq. (\ref{eq:du-do}) we have
\begin{eqnarray}
\omega^{\rho}\partial_{\rho}u_{\sigma} & = & \frac{1}{2}(u\cdot\omega)\dot{u}_{\sigma},  \nonumber \\
\omega^{\sigma}\partial_{\rho}u_{\sigma} & = & \frac{1}{2}(u\cdot\omega)\dot{u}_{\rho},  \nonumber \\
u_{\sigma}\partial_{\rho}\omega^{\sigma} & = & \frac{3}{2}(u\cdot\omega)
\dot{u}_{\sigma}.  
\label{eq:omega-du}
\end{eqnarray}

Then we can derive identities for $B_{\alpha}$ and $E_{\alpha}$. From 
$E^{\rho}=\epsilon^{\rho\alpha\beta}u_{\alpha}B_{\beta}$ we can easily see 
$E\cdot B=0$. Using $\partial_{\sigma}(\beta\mu)=\beta QE_{\sigma}$, 
$E^{\rho}=\epsilon^{\rho\alpha\beta}u_{\alpha}B_{\beta}$ and 
$\dot{u}^{\alpha}=\partial^{\alpha}\ln T$, we can derive
\begin{equation}
(E\cdot\omega)u^{\lambda}  =  B\cdot\partial u^{\lambda},
\end{equation}
which leads to
\begin{equation}
E\cdot\omega=B\cdot\partial u^{\lambda}=\dot{u}_{\beta}B^{\beta}
=u_{\rho}\partial^{\rho}(u\cdot B)=0.
\end{equation}

\section{Proof of $G=0$}

\label{sec:G=0}
Let us fix $G$ in $\mathscr{V}_{1}^{\mu}$ under some physical constraints.
As we have shown in Eq. (\ref{eq:ordering}), the Wigner operator
has the medium and vacuum part. We obtain the medium part
involving $\mathscr{V}^{\mu}_1$ by solving Eq.\ (\ref{eq-c}). 
If we take $V\rightarrow 0$, i.e. there are no particles in the
system, the medium part should vanish. Therefore, $G$ must be a function of $V$, 
its derivative $V_{u\cdot p}^{\prime}$ and $C$. Simply, we can express $G$
as polynomials of $V$, $V_{u\cdot p}^{\prime}$ and $C$. 

In the framework of the Boltzmann equation, 
the distribution function $f$ can be expanded near equilibrium, 
\begin{equation}
f=f_{0}+f_{1}+...,
\end{equation}
where $f_{0}$ is the distribution function in equilibrium and the first order correction, 
$f_{1}=-(\text{dissipative terms})\times T\frac{\partial}{\partial(u\cdot p)}f_{0}$, 
is linear in $f_{0}$. In our case, $V$, $V_{u\cdot p}^{\prime}$ and $C$
(derivatives of $V$) correspond to $f_{0}$ and $T\frac{\partial}
{\partial(u\cdot p)}f_{0}$. Therefore, we can assume $G$ is 
also a linear combination of $V$, $V_{u\cdot p}^{\prime}$ and $C$.

Including all possible contractions of the vectors $u^{\mu}$, $\omega^{\mu}$, 
$B^{\mu}$ and $p^{\mu}$ of the first order, we then have the following form for $G$, 
\begin{eqnarray}
G &=& V \left[ \sum_{i=0} \frac{1}{m^{i+2}} (p\cdot\omega) (u\cdot p)^i X_{1,i} 
+\sum_{i=0} \frac{1}{m^{i+1}}(u\cdot\omega)(u\cdot p)^{i}X_{2,i}\right. \nonumber\\
&&\left. +\sum_{i=0} \frac{1}{m^{i+3}}(p\cdot B) (u\cdot p)^i X_{3,i}
+\sum_{i=0} \frac{1}{m^{i+2}}(u\cdot B)(u\cdot p)^{i}X_{4,i}
\right] \nonumber\\
&&+V_{u\cdot p}^{\prime} \left[ \sum_{i=0} \frac{1}{m^{i+1}} (p\cdot\omega) (u\cdot p)^i Y_{1,i} 
+\sum_{i=0} \frac{1}{m^{i}}(u\cdot\omega)(u\cdot p)^{i}Y_{2,i}\right. \nonumber\\
&&\left. +\sum_{i=0} \frac{1}{m^{i+2}}(p\cdot B) (u\cdot p)^i Y_{3,i}
+\sum_{i=0} \frac{1}{m^{i+1}}(u\cdot B)(u\cdot p)^{i}Y_{4,i} \right] \nonumber\\
&&+C \left[ \sum_{i=0} \frac{1}{m^{i+1}} (p\cdot\omega) (u\cdot p)^i Z_{1,i} 
+\sum_{i=0} \frac{1}{m^{i}}(u\cdot\omega)(u\cdot p)^{i}Z_{2,i}\right. \nonumber\\
&&\left. +\sum_{i=0} \frac{1}{m^{i+2}}(p\cdot B) (u\cdot p)^i Z_{3,i}
+\sum_{i=0} \frac{1}{m^{i+1}}(u\cdot B)(u\cdot p)^{i}Z_{4,i} \right]  , 
\end{eqnarray}
where $X_{j,i}$, $Y_{j,i}$, $Z_{j,i}$ are dimensionless constants and
all dependence on $\mu$ and $T$ are through $V$, $V_{u\cdot p}^{\prime}$ and $C$. 
From Eq.\ (\ref{eq:V_1_temp_02},\ref{eq:F_01}), we neglect
other complicated expressions, e.g. $\log (u\cdot p/m)$. 
We also assume that macroscopic quantities should not appear in denominators, 
e.g. terms like $1/(u\cdot B)$ will be divergent at vanishing magnetic field and should 
be absent in our discussions. 

Since $C\delta(p^{2}-m^{2})\propto\delta(u\cdot p\pm m)\delta(\mathbf{p})$, 
the $Z_{1,i}$ and $Z_{3,i}$ terms vanish, and only $Z_{2,0}$ and $Z_{4,0}$ terms 
survive. Although we consider a system of massive fermions, we would not expect 
any divergences in the current when we take the massless limit, i.e. 
$\int d^{3}p p^{\mu} G \delta(p^{2}-m^{2})$ should be finite. 
Then all $1/m^i$ terms with $i > 0$ should be gone.  Then $G$ becomes,
\begin{equation}
G=C(u\cdot\omega)Z_{2,0}+C\frac{1}{m}(u\cdot B)Z_{4,0}.
\end{equation}

We have already proved that all terms in $\mathscr{V}_{1}^{\mu}$ except $G$ satisfy 
the energy-momentum conservation, $\partial_{\mu}T^{\mu \nu}=F^{\nu\lambda}j_{\lambda}$. 
Therefore $G$ has to satisfy it separately, but it does not do automatically unless $Z_{2,0}$ 
and $Z_{4,0}$ vanish. Finally we reach $G=0$.  

\bibliography{reference-list}

\end{document}